\documentclass{svproc}
\usepackage{graphicx}
\usepackage{caption}
\usepackage{subcaption}
\usepackage{dcolumn}
\usepackage{bm}
\usepackage{multirow}
\usepackage{color}
\usepackage{amsmath}
\usepackage{textcomp}
\usepackage{amssymb}
\usepackage{hyperref}
\usepackage{dcolumn}
\usepackage{color}
\usepackage[normalem]{ulem}
\usepackage{soul,xcolor}

\newcommand{\nuc}[2]{$^{{#1}}${#2}}

\newcommand{\be}{\begin{equation}}
\newcommand{\ee}{\end{equation}}

\newcommand{\bea}{\begin{eqnarray}}
\newcommand{\eea}{\end{eqnarray}}

\usepackage{url}

\begin{document}
\mainmatter              
\title{Constraining level densities using spectral data}
\titlerunning{Level densities and spectra}  
%
\author{G. P. A. Nobre\inst{1} \and D. A. Brown \inst{1} \and M. W. Herman \inst{1,2} }
\authorrunning{G.P.A. Nobre et al.} 
%
\tocauthor{Gustavo P. A. Nobre, David A. Brown, Michal W. Herman}
\institute{National Nuclear Data Center, Brookhaven National Laboratory, Upton, NY 11973-5000, USA,\\
\email{gnobre@bnl.gov},\\ 
\and
Los Alamos National Laboratory, Los Alamos, NM 87545, USA}

\maketitle              

\begin{abstract}
Several models of level densities exist and they often make simplified assumptions regarding the overall behavior of the total level densities (LD) and the intrinsic spin and parity distributions of the excited states. Normally, such LD models are constrained only by the measured $D_0$, i.e. the density of levels at the neutron separation energy of the compound nucleus (target plus neutron), and the sometimes subjective extrapolation of discrete levels. In this work we use microscopic Hartree-Fock-Bogoliubov (HFB) level densities, which intrinsically provide more realistic spin and parity distributions, and associate variations predicted by the HFB model with the observed double-differential cross sections at low outgoing neutron energy, region that is dominated by the LD input. With this approach we are able to perform fits of the LD based on actual experimental data, constraining the model and ensuring its consistency. This approach can be particularly useful in extrapolating the LD to nuclei for which high-excited discrete levels and/or values of $D_0$ are unknown. It also predicts inelastic gamma (n,n$^{\prime}\gamma$) cross sections that in some cases can differ significantly from more standard LD models such as Gilbert-Cameron.
\end{abstract}
\setstcolor{red}

\section{Introduction}

As the nuclear excitation energy grows, due to the exponential increase of the number of levels, one must deal with the level densities (LD) rather than with individual levels.
Several phenomenological models exist to describe the general behavior of (LD), such as the Gilbert-Cameron (GC) \cite{GC} and others, which assume  simplified functional forms of the LD and  are constrained by the often limited availability of experimental data. It is known that there are only a few ways to experimentally constrain LD, such as through the $D_0$ (LD at neutron separation energy of the compound nucleus) or the matching at the excitation energy region transitioning from discrete levels to LD.


More fundamental and predictive LD models like the microscopic Hartree-Fock-Bogoliubov (HFB)  \cite{HFBM} incorporated to the RIPL-3 parameter library \cite{RIPL-3} provide more global and consistent  LD, based on the intrinsic structure properties of nuclei and observed distribution of discrete levels. This brings reliability to the LD in the whole range of excitation energy, not only near the discrete-level cutoff or at $D_0$. Additionally, the HFB model provides more realistic spin and parity distributions which emerge naturally from the model, while a phenomenological model such as GC simplistically assumes equal distributions.

In nuclear reaction evaluations, however, predictive models are seldom employed since the greater flexibility of parameter fitting of phenomenological models may lead to better cross section agreements \cite{WONDER2015}. In this work we aim to circumvent this apparent deficiency of the HFB LD model by using experimental data from neutron double-differential spectra cross sections on \nuc{56}{Fe} to impose direct constraints on the HFB total LD for \nuc{56}{Fe} and \nuc{56}{Mn}. We show that we can obtain a more realistic LD and at least equally good cross sections compared to a fitted Gilbert-Cameron model, in particular for the \nuc{56}{Fe}(n,p) reaction which is of dosimetry interest. This way we can combine the predictive power of a microscopic model with good description of observed data, as required by a variety of applications. 
We also investigate the consequences of this approach in the prediction of inelastic-gamma cross sections 
as compared to measured  data.

\section{Description of LD models}

LD models are crucial for Hauser-Feshbach and pre-equilibrium reaction mechanisms. Phenomenological models tend to better reproduce average behaviors while missing detailed structure components. We will discuss the phenomenological GC and the microscopic HFB models, as 
implemented 
in 
EMPIRE \cite{EMPIRE}. 

\subsection{Gilbert-Cameron model}

Most phenomenological LD models are based in some form on the analytical expression derived from the Fermi Gas Model \cite{GC}. We assume that the  density of intrinsic levels with spin $J$, parity $\pi$ and excitation energy $E_x$ can be factored  in terms of its state density and spin and parity dependence.
The Gilbert-Cameron model \cite{GC} splits the excitation energy range in two parts, with different functional forms applied to each of them. Below a chosen matching energy $U_x$ a constant temperature state density is employed while above $U_x$ the Fermi Gas state density is adopted, with pairing energy given by $\Delta = n\tfrac{12}{\sqrt{A}} $, where $A$ is the nucleus mass number and $n$ is 0, 1, or 2 for odd-odd, odd-even, and even-even nuclei, respectively. Some model parameters 
are internally determined by  imposing that the total LD and its derivative are continuous at the matching point $U_x$.


\subsection{HFB model}
\label{sec:HFB-model}

EMPIRE has implemented within its options the microscopic combinatorial approach \cite{HFBM} developed for RIPL-3 \cite{RIPL-3}. It consists of using single-particle level schemes obtained from constrained axially symmetric Hartree-Fock-Bogoliubov method (HFBM) based on the BSk14 Skyrme force 
to construct incoherent particle-hole state densities 
as functions of the excitation energy $E_x$, the spin projection $M$ (on the intrinsic symmetry axis of the nucleus) and the parity $\pi$.
Collective effects are incorporated through a boson partition function 
providing vibrational state densities dependent on multipolar phonon energies.

\section{Implementing constraints from neutron differential spectra data}
\label{sec:xsec}

\begin{figure}
\includegraphics[scale=0.50,keepaspectratio=true,clip=true,trim=2mm 0mm 5mm 0mm]{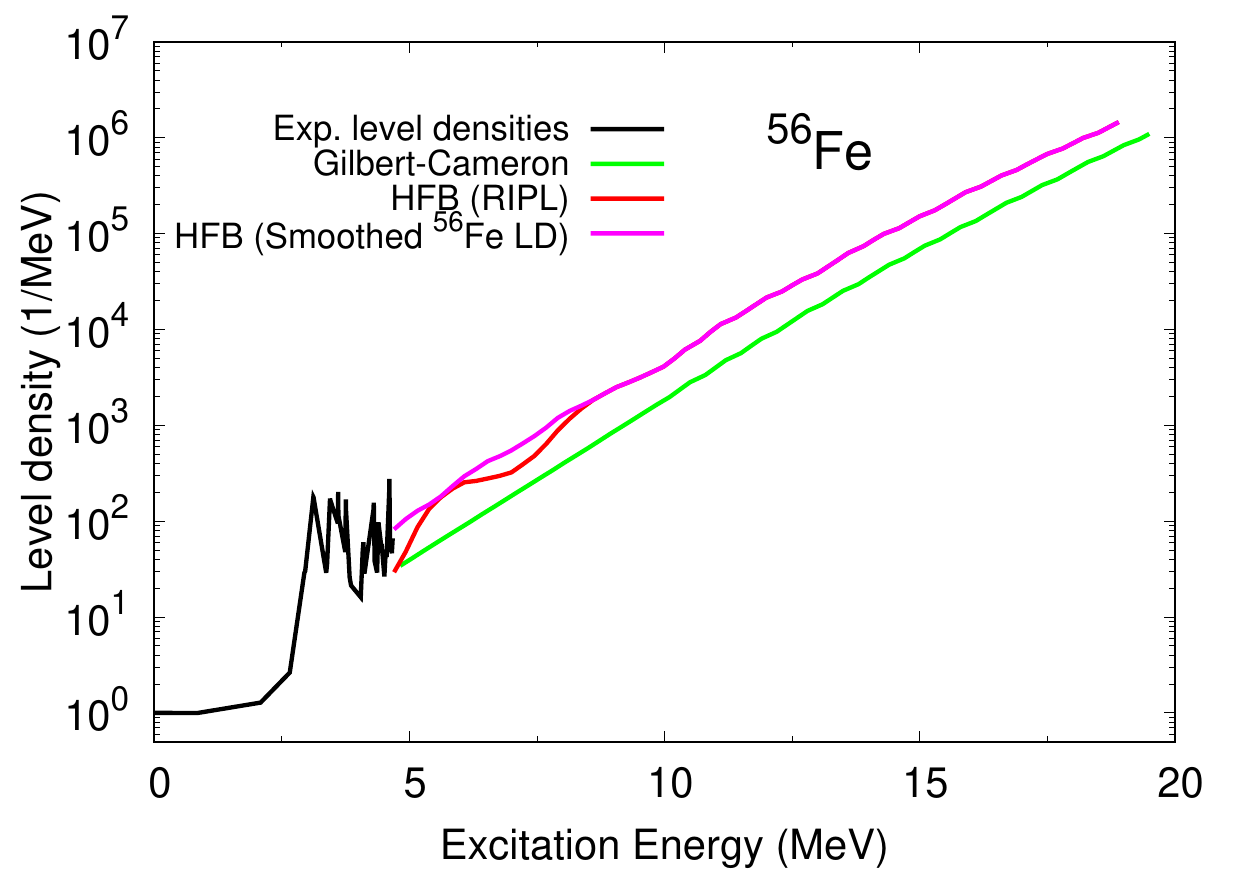}
 \includegraphics[scale=0.50,keepaspectratio=true,clip=true,trim=2mm 0mm 5mm 0mm]{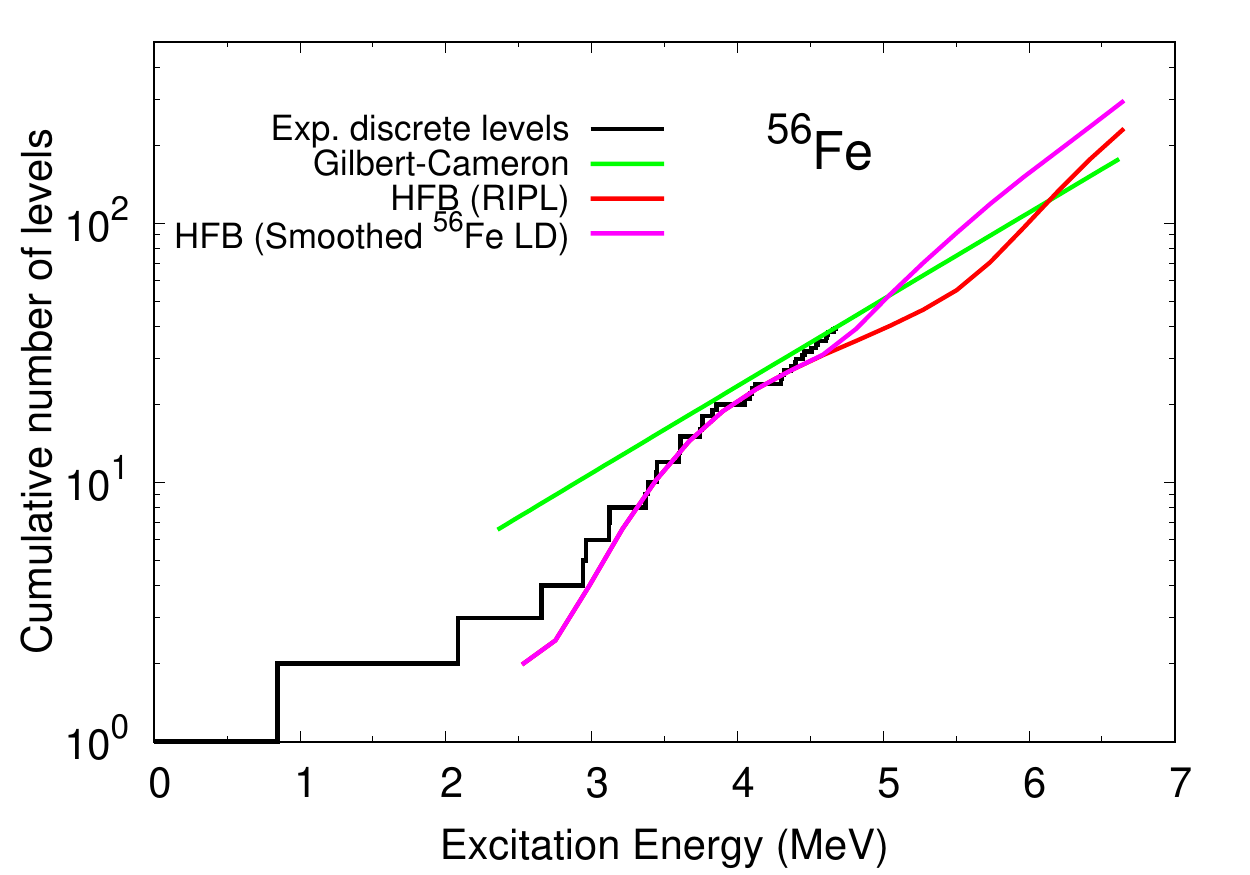}
 \caption{Level densities and cumulative number of levels of \nuc{56}{Fe} for the different LD approaches explained in Section~\ref{sec:xsec}.}\label{fig:fe56-ld}
 \end{figure}
\begin{figure}[t]
 \includegraphics[scale=0.50,keepaspectratio=true,clip=true,trim=2mm 0mm 5mm 0mm]{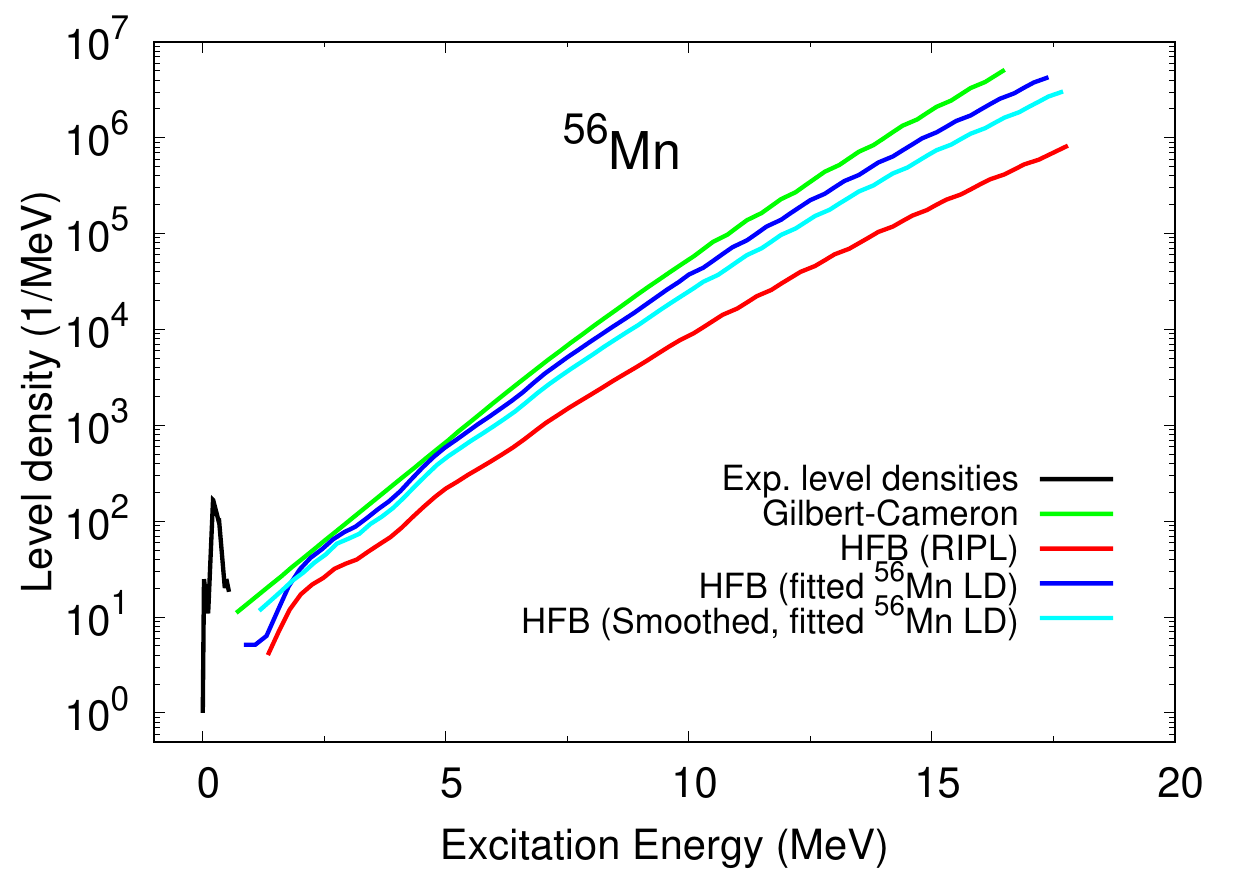}
 \includegraphics[scale=0.50,keepaspectratio=true,clip=true,trim=2mm 0mm 5mm 0mm]{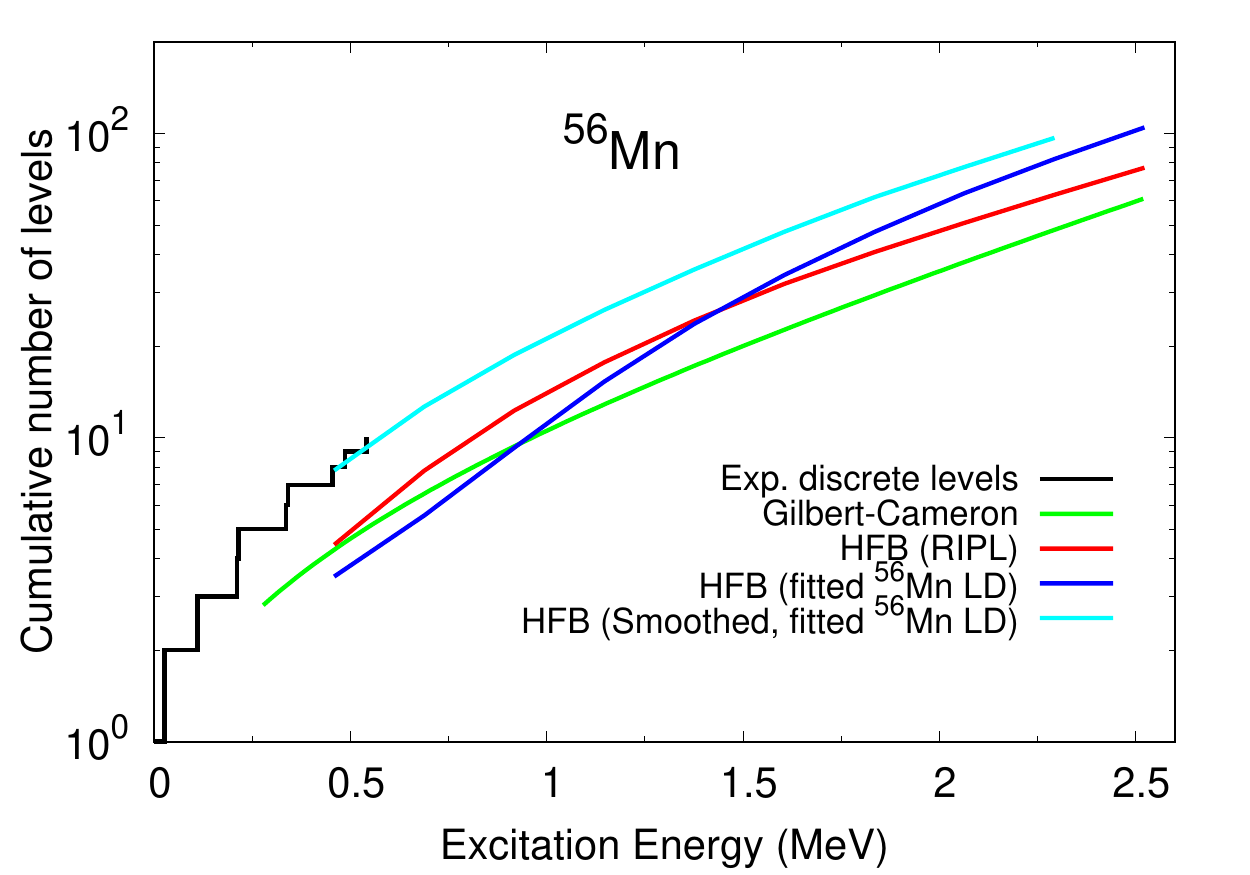}
 \caption{Level densities and cumulative number of levels of \nuc{56}{Mn} for the different LD approaches explained in Section~\ref{sec:xsec}.}\label{fig:mn56-ld}
 \end{figure}

We adopted the $n$ + \nuc{56}{Fe} reaction as our test case to identify the impact of details of LD in the cross sections, using the same parametrization employed in the fast-region evaluation of \nuc{56}{Fe} present in the ENDF/B-VIII.0 evaluation \cite{CIELO-IRON,ENDF-VIII.0} as part of the CIELO project \cite{CIELO}, ensuring that all calculated cross sections are mutually consistent and in good agreement with experimental data. 
We performed reaction calculations using five different approaches for the LD: 
\emph{a)} assuming the GC model for all nuclei, as done in the \nuc{56}{Fe} evaluation \cite{CIELO-IRON,ENDF-VIII.0} (green curves in Figures~\ref{fig:fe56-ld}-\ref{fig:ddx}); 
\emph{b)} assuming HFB LD as available from RIPL, with no modifications to it (red curves in Figures~\ref{fig:fe56-ld}-\ref{fig:fe56-np}); 
\emph{c)} same as \emph{b)} but fitting two parameters of \nuc{56}{Mn} LD to (n,p) data (blue curves in Figures~\ref{fig:mn56-ld}-\ref{fig:ddx}); 
\emph{d)} same as \emph{c)} but with the \nuc{56}{Fe} LD structures smoothed out in order to improve agreement with experimental data of neutron double-differential spectra (magenta curves in Figures~\ref{fig:fe56-ld},~\ref{fig:fe56-np} and~\ref{fig:ddx});
\emph{e)} same as \emph{d} but with \nuc{56}{Mn} smoothed and refitted to (n,p) cross sections (cyan curves in Figures~\ref{fig:mn56-ld} and~\ref{fig:fe56-np}).


\section{Discussion}
\label{sec:discussion}

\begin{figure}[h]
 \centering
 \includegraphics[scale=0.70,keepaspectratio=true,clip=true,trim=1mm 0mm 4mm 0mm]{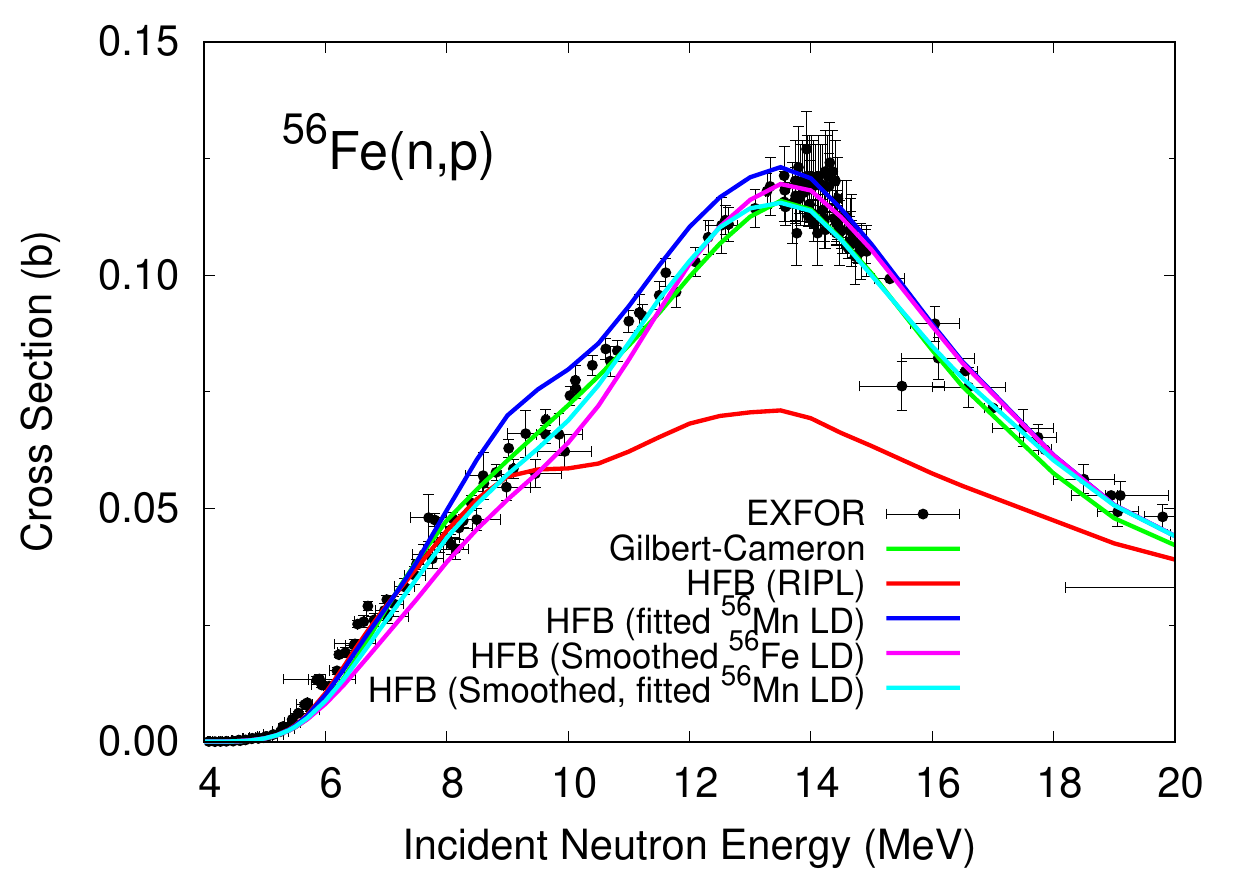}
 \caption{\nuc{56}{Fe}(n,p)\nuc{56}{Mn} cross section obtained from the adoption of the different LD approaches explained in Section~\ref{sec:xsec}. Experimental data from EXFOR \cite{EXFOR}.}\label{fig:fe56-np}
\end{figure}

\begin{figure}[h]
 \centering
 \includegraphics[scale=0.51,keepaspectratio=true,clip=true,trim=0mm 0mm 5mm 0mm]{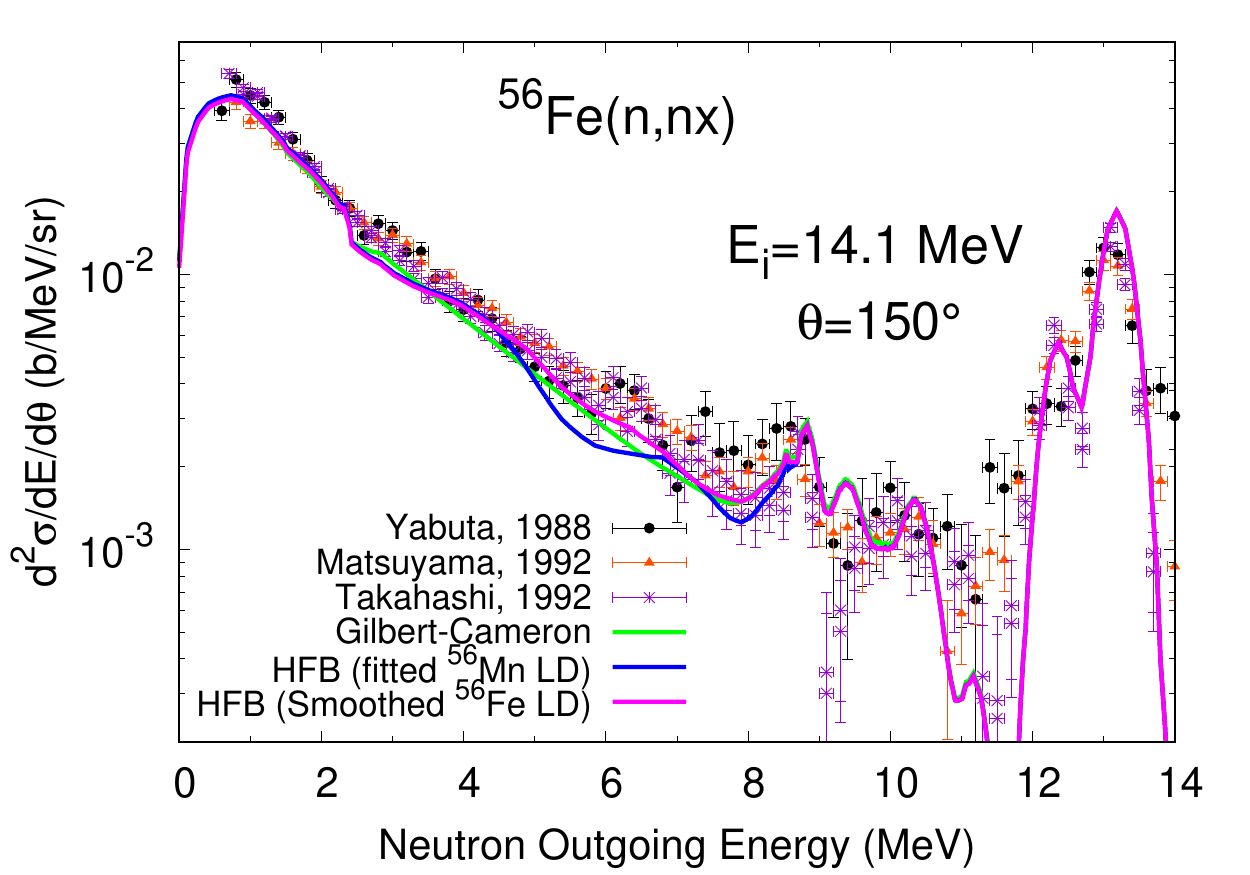} 
%
 \includegraphics[scale=0.51,keepaspectratio=true,clip=true,trim=7mm 0mm 5mm 0mm]{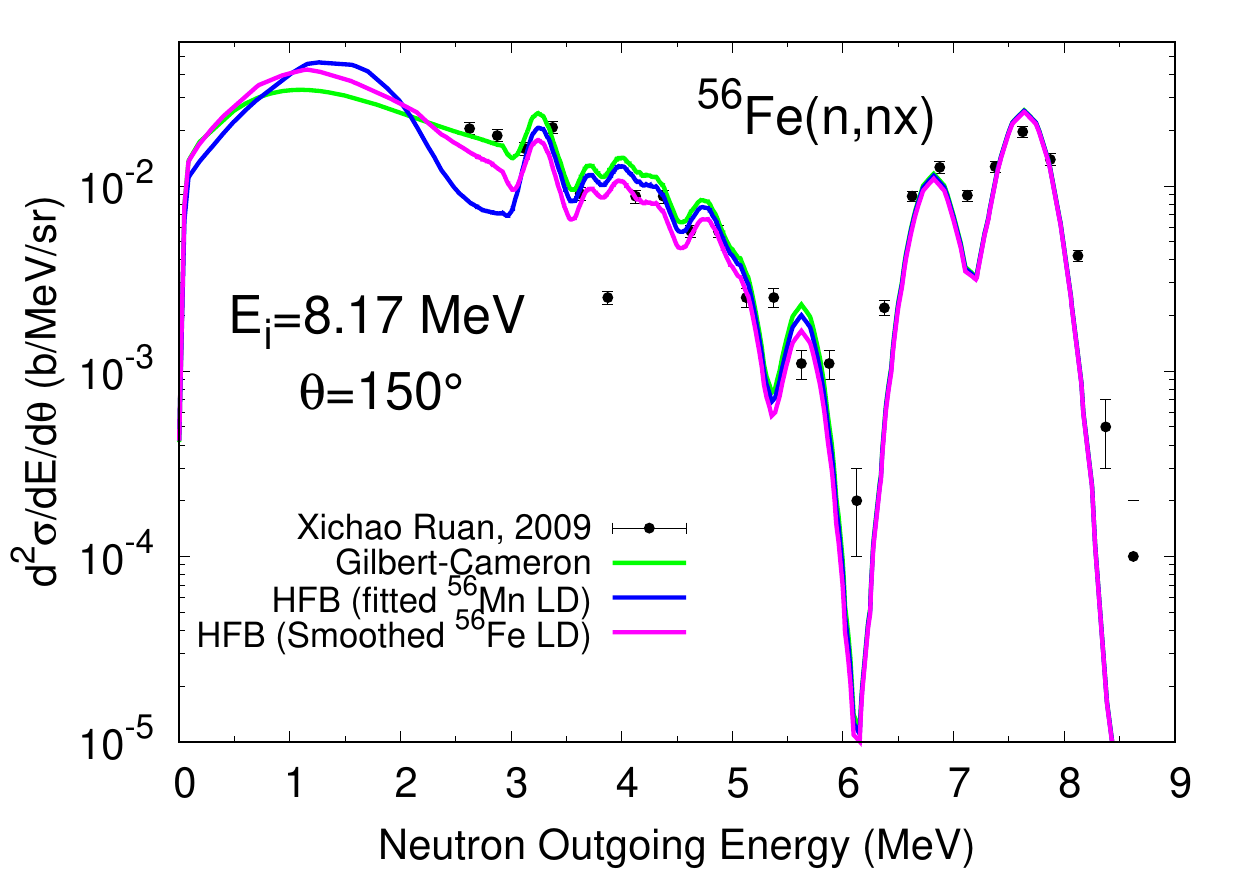}
 \caption{Example of double-differential spectra for at 150 degrees and neutron incident energies of 14.1 MeV (left panel) and 8.17 MeV (right panel) for the different LD approaches explained in Section~\ref{sec:xsec}. Data from EXFOR \cite{EXFOR}.}\label{fig:ddx}
\end{figure}

By comparing the green and red curves in Figure~\ref{fig:fe56-ld} we see that while the GC LD is smooth (as it comes from constant-temperature analytical forms), the HFB LD present fluctuations, or structures, in the range  $5 \lesssim E_x \lesssim 9$~MeV. Both GC and HFB (from RIPL) models approximately reproduce reasonably well the number of levels at around 4.5~MeV which is around where one would normally impose the transition from the discrete levels to LD. This transition point, or excitation energy cut-off, can however be rather arbitrary. One can clearly see from Figure~\ref{fig:fe56-ld} that the HFB predicted cumulative number of levels yields a much better agreement with the overall behavior of observed discrete levels, which makes it much more independent from the choice of excitation energy at which the transition to LD is made. Even though these two apparently similarly-reasonable (from the perspective of discrete-level matching) LD models, they lead to dramatically different (n,p) cross sections (Figure~\ref{fig:fe56-np}). Even after fitting \nuc{56}{Mn} LD parameters (blue curve), the agreement with (n,p) data is still not optimal. Therefore, additional constraints for the LD are needed. By noticing direct correlations between the \nuc{56}{Fe} LD for a given $E_x$ region and the DD cross section at certain neutron-outgoing energies, we were able to use the experimental knowledge of DD spectra to impose constraints on LD. For this we smoothed the structures of the HFB LD to the point that effects of these structures would not appear in calculated DD spectra and that the agreement with DD experimental data would be satisfactory (Figure~\ref{fig:ddx}).   The result of this is shown as the magenta curves. Even though this produced a considerably better agreement with (n,p) data, this is still not as good as the GC one. This can be remediated by smoothing and refitting the \nuc{56}{Mn} LD to minimize $\chi^2$ relative to (n,p) experimental data. This resulted in the cyan curves. In addition to obtaining better (n,p) cross sections, this also leads to a more realistic \nuc{56}{Mn} LD relative to the observed discrete levels (Figure~\ref{fig:mn56-ld}).

\section{Impact on inelastic gammas}
\label{sec:inel-gammas}

Another application of using experimental DD spectra to constrain HFB level densities is in the description of inelastic gamma cross-section data. Recently, cross-section measurements of gamma emissions corresponding to transitions between excited levels have provided new information which is very useful to complement neutron and reaction cross sections in usual neutron evaluations. From a theoretical standpoint, predicting and consistently fitting gamma cross sections can be a challenge due to the variety of mechanisms involved. Therefore, a more predictive and fundamental LD model would provide better reliability for calculated gamma cross sections. We have done the comparison between GC and modified HFB models for all transitions measured in the work of Negret et al. \cite{Negret}, and also other transitions that were not measured. We have found that in some cases there are very little differences. However, for some transitions there are noticeable differences in calculated cross sections, with the HFB one generally aligning better with observed data. Much greater discrepancy was observed in transitions involving states of opposite parity of that of the ground state, which gives more significance to a more predictive and internally consistent set of LD, especially when experimental data are not available.

\section{Summary and conclusions}
\label{Sec:Conclusion}

Even though it is known that cross sections strongly depend on level densities (LD) there are normally very little direct experimental input in the determination of their details. In this work we explored this feature by imposing constraints in extended regions of LD by observing their impact on the agreement of neutron double-differential spectra with experimental data. This allowed us to extract experiment-based information about LD that is useful for the structure theory community which develop microscopic LD models, as well as to increase the internal self-consistency of models used reaction calculations leading to evaluation-level quality of cross-sections. Additionally, this proved to be a method to obtain more reliable inelastic gamma cross sections, in particular for those transitions without measured data or those involving levels with parity opposite of that of the ground state. One might expect that these effects are more pronounced for the nuclei close to the shell closures.

\section{Acknowledgments}

The work at Brookhaven National Laboratory was sponsored by the Office of Nuclear
Physics, Office of Science of the U.S. Department of
Energy under Contract No. DE-AC02-98CH10886 with
Brookhaven Science Associates, LLC. Work at Los Alamos National Laboratory was carried out under the auspices of the National Nuclear Security Agency of the U.S. Department of Energy under Contract No. DE-AC52- 06NA25396.

%
%


%

%





\end{document}